\documentclass[twocolumn]{aastex62}

\def\lapp{\ifmmode\stackrel{<}{_{\sim}}\else$\stackrel{<}{_{\sim}}$\fi}
\def\gapp{\ifmmode\stackrel{>}{_{\sim}}\else$\stackrel{>}{_{\sim}}$\fi}
\usepackage{multirow}
\usepackage{color}
\usepackage{amsmath}
\usepackage{soul}
\usepackage{hyperref}

\shorttitle{}
\shortauthors{}
\begin{document}

\title{{\it NuSTAR} hard X-ray studies of the pulsar wind nebula 3C~58}

\correspondingauthor{Hongjun An}
\email{hjan@cbnu.ac.kr}

\author{Hongjun An}
\affiliation{Department of Astronomy and Space Science,\\
Chungbuk National University, Cheongju, 28644, Republic of Korea}


\begin{abstract}
        We report on new {\it NuSTAR} and archival {\it Chandra} observations of
the pulsar wind nebula (PWN) 3C~58.
Using the X-ray data, we measure energy-dependent morphologies
and spatially-resolved spectra of the PWN. We find that the PWN size becomes smaller with
increasing energy and that the spectrum is softer in outer regions.
In the spatially integrated spectrum of the PWN, we find a hint of a spectral break at $\sim$25~keV.
We interpret these findings using synchrotron-radiation scenarios.
We attribute the size change to the synchrotron burn-off effect.
The radial profile of the spectral index has a break at $R\sim80''$,
implying a maximum electron energy of $\sim$200\,TeV which is larger than a previous estimate,
and the 25-keV spectral break corresponds to a maximum electron energy of $\sim$140\,TeV
for an assumed magnetic field strength of 80\,$\mu$G. Combining the X-ray data and a previous radio-to-IR SED,
we measure a cooling break frequency to be $\sim 10^{15}$\,Hz,
which constrains the magnetic-field strength in 3C~58 to be 30--200$\mu$G for an assumed
age range of 800--5000 years.
\end{abstract}

\keywords{ISM: individual (3C~58) --- ISM: jets and outflows --- X-rays: ISM --- stars: winds, outflows}

\section{Introduction}
\label{sec:intro}
	Pulsar wind particles are believed to be accelerated to very high energies
in the termination shock of pulsar wind nebulae \citep[PWNe; e.g.,][]{kc84a}, and the energetic particles
emit radiation across the whole electromagnetic waveband as observed in
broadband spectral energy distributions (SEDs) of young PWNe.
The emission is often modeled with synchrotron and inverse-Compton radiation \citep[e.g.,][]{baa11}, where
interaction of particles and magnetic fields produces synchrotron emission in the radio
to X-ray band, and the particles upscatter soft photons (e.g., CMB, IR and optical backgrounds)
to produce high-energy GeV to TeV photons \citep[e.g.,][]{3c58magic14}. In this scenario, the X-ray SED is
particularly important as it is a direct imprint of the particle distribution
at the highest-energy range \citep[e.g.,][]{kkcp17}, and the maximum particle energy can
help to understand particle acceleration in relativistic shocks of PWNe \citep[e.g.,][]{skl15}.
Particle distributions in PWNe were inferred from their photon spectra for some young PWNe, but
high-energy cutoffs in the photon spectra corresponding to the maximum electron energies
were measured only for a few sources.

	As particles flow in a PWN, the particles lose their energy to radiation and
become less energetic with distance from the central pulsar.
This effect can be seen in morphologies, spectral
maps \citep[e.g.,][]{amrk+14,kkcp17}, and spatially-integrated spectra \citep[e.g.,][]{nhra+14,mrha+15}
in the X-ray band. A change of the morphology can tell us about evolution of magnetic field
and particle density \citep[e.g.,][]{r09}, and spatial variation of the spectrum can be used to
study particle flow in PWNe \citep[e.g.,][]{tc12}. These studies have been conducted for
X-ray bright PWNe \citep[see][for a review]{rpkk+17}.

	The TeV-detected PWN 3C~58 powered
by a 65-ms pulsar \citep[PSR~J0205+6449;][]{mssr+02}
is a young object possibly associated with SN~1181 \citep[][]{s1971}.
The association is not very firm as speed of radio expansion \citep[][]{bkw01} or optical knots \citep{f1983}
suggests that the age may be larger, perhaps close to the pulsar spin-down age of 5000\,yr.
The distance to the PWN is estimated to be 2\,kpc \citep[][]{k2013}
or 3\,kpc \citep[][]{rgkh+93} with H~I observations. The distance and age inferred from observations
may be somewhat correlated \citep[][]{k2013}, and an independent estimation of the age
can be made with X-ray SED measurements and time-evolving PWN models \citep[e.g.,][]{shrg+08}.

	Because 3C~58 is an intriguing PWN with torus-jet structure,
it was intensively studied in the soft ($<8$\,keV) X-ray band
\citep[e.g.,][]{tskh+00,bwml+01,shvm04}. Although its soft X-ray spectrum is well modeled with
a simple power law, it is important to see if the spectrum extends to
hard X-ray band above 10\,keV. This can be done with {\it NuSTAR} \citep[][]{hcc+13} and may give us new
insights into particle acceleration and transport in PWNe.
Although {\it NuSTAR} angular resolution is not sufficient to
separate the pulsar from the PWN \citep[][]{amwb+14}, the
low beaming fraction of PSR~J0205+6449 \citep[$<$0.2;][]{lrck+09,khud+10}
allows temporal separation of the two emission components, and hence hard X-ray
properties of the PWN can be well studied with {\it NuSTAR}.

	Here we report on X-ray observations of 3C~58 made with {\it NuSTAR} and {\it Chandra}.
We measure energy-dependent morphology, spatial variation of the spectral index, and a spatially integrated
broadband X-ray spectrum. These measurements are used to infer properties of 3C~58 with
synchrotron-radiation scenarios. We present analysis results in Sections~\ref{sec:sec2} and construct
a broadband SED in the radio to X-ray band in Section~\ref{sec:sec3}.
We then discuss and conclude in Section~\ref{sec:sec4}.

\section{Observational Data and Analysis}
\label{sec:sec2}

\subsection{Data reduction}
\label{sec:sec2_1}
\newcommand{\markaa}{\tablenotemark{a}}
\begin{table}[t]
\vspace{-0.0in}
\begin{center}
\caption{Summary of X-ray observations used in this work
\label{ta:ta1}}
\vspace{-0.05in}
\scriptsize{
\begin{tabular}{ccccc} \hline\hline
Observatory       &  Obs. ID    &  Instruments    &   Exposure   \\
                  &             &                 &     (ks)     \\ \hline
{\it Chandra}     &    1848     &     HRC-S       &  33          \\
{\it Chandra}     &    2756     &     HRC-S       &  24          \\
{\it Chandra}     &    3832     &     ACIS-S      &  136         \\
{\it Chandra}     &    4382     &     ACIS-S      &  167         \\
{\it Chandra}     &    4383     &     ACIS-S      &  39          \\
{\it NuSTAR}      & 30301011001 &     FPMA,B      &  81          \\ \hline
\end{tabular}}
\end{center}
\vspace{-0.5 mm}
\end{table}

       We use new {\it NuSTAR} and archival {\it Chandra} data shown in Table~\ref{ta:ta1}.
We process the data using {\tt nustardas} integrated in HEASOFT~6.24 and {\tt chandra\_repro}
of CIAO~4.10 along with the most recent calibration data files.
Note that we use strict filters ({\tt saamode=optimized} and {\tt tentacle=yes})
for the {\it NuSTAR} data process to reduce background around South Atlantic Anomaly (SAA) passage.

\subsection{X-ray timing analysis}
\label{sec:sec2_2}
	X-ray and gamma-ray timing studies of the pulsar PSR~J0205+6449 have been
done previously \citep[e.g.,][]{lrck+09,khud+10,3c58LAT09,fermi2PC,ltlg+18}.
Here we perform pulsar timing analyses to select off-pulse intervals for our PWN study.
The {\it Chandra} HRC data \citep[][]{shm02} are particularly useful for estimating the pulsar's
constant emission which may contaminate the PWN spectrum (Sections~\ref{sec:sec2_4} and \ref{sec:sec2_5}).
So we reanalyze them here.

\begin{figure}
\centering
\hspace{0 mm}
\includegraphics[width=3.35 in]{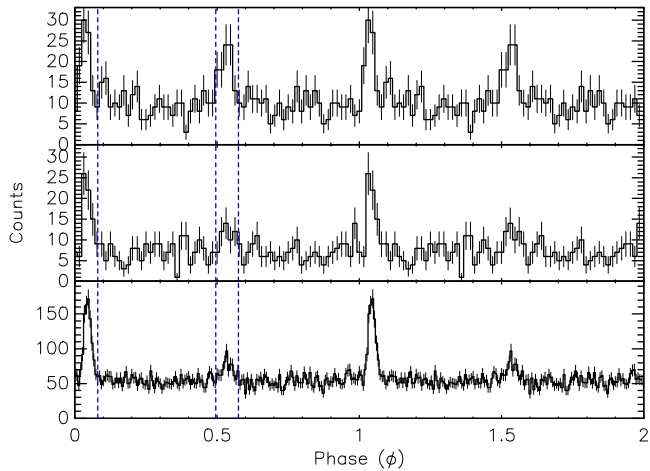} 
\figcaption{X-ray pulse profiles measured with {\it Chandra} (top and middle) and {\it NuSTAR} (bottom).
Vertical blue lines show on-pulse intervals.
Pulsations are detected with high significances: $H$=57, 33, and 599 from top to bottom.
For the {\it Chandra} profiles, we use an $R=1''$ region with the PIs
30--240 in Obs. IDs 1848 (top) and 2756 (middle). An $R=30''$ circle and the 3--79\,keV band are used
for the {\it NuSTAR} profile.
\label{fig:fig1}
}
\vspace{0mm}
\end{figure}

	For a {\it Chandra} HRC data analysis, we extract events within an $R=1''$ circular region
with PI channels between 30 and 240. We then search for pulsation around
the reported period of 65\,ms using the $H$ test \citep[][]{drs89}, and find
pulsations in both HRC data sets (Fig.~\ref{fig:fig1}).
We also perform a pulsation search in the {\it NuSTAR} data using an $R=30''$ circle
in the 3--79\,keV band and find pulsations (Fig.~\ref{fig:fig1}).
Because of very hard pulsed emission of the pulsar and {\it NuSTAR}'s excellent sensitivity
at high energies \citep[][]{mhma+15}, the source pulsations are detected easily in a
short $<$10-ks interval. So we are able to perform a phase semi-coherent timing
analysis using a double Gaussian function for the pulse profile.
The data are sensitive enough to measure the
first frequency derivative; the measured frequency and its derivative
are 15.20102356(11)$\rm \ s^{-1}$ and
$-4.44(62)\times10^{-11}\ \rm s^{-2}$ at the epoch MJD~58252.479.
Results of pulsar timing and spectral analyses will be presented elsewhere.
We show the measured pulse profiles in Figure~\ref{fig:fig1} and select off-pulse
intervals ($\phi$=0.08--0.495 and 0.575--1.0) for the PWN studies below.
Note that changing the off-pulse intervals slightly (e.g., $\phi$=0.1--0.49 and 0.59--1.0)
does not alter the results below.

\subsection{X-ray image analysis}
\label{sec:sec2_3}
	Because {\it NuSTAR} angular resolution is not sufficient to separate
the pulsar from the PWN spatially, the pulsar's constant emission may contaminate the PWN spectrum.
This contamination can be estimated with high-resolution {\it Chandra} HRC data
\citep[e.g., PSR~B1509$-$58;][]{cakh+16}, and so we use {\it Chandra} HRC observations
to estimate the pulsar contamination in 3C~58.

\begin{figure}
\centering
\includegraphics[width=3.3 in]{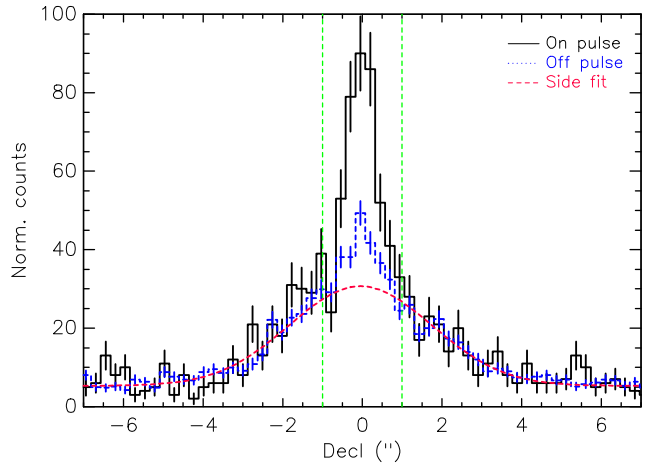}
\figcaption{Time-normalized brightness profiles in the 0.3--5\,keV band measured with the HRC data.
The black and the blue lines show profiles for the on-pulse and the off-pulse intervals,
respectively. The red dashed line denotes a fit to the $|D|>1''$ off-pulse profile, representing
the PWN emission. Vertical dashed lines (green) mark a central 2$''$ region in which
$>$95\% pulsar emission is contained.
\label{fig:fig2}
}
\vspace{0mm}
\end{figure}

	We combine the HRC observations (Table~\ref{ta:ta1}) after aligning the images
using the pulsar position and select on- and off-pulse intervals.
The on-pulse intervals are shown in Figure~\ref{fig:fig1}.
We use a rectangular region ($30''\times 30''$) to include the torus-jet structure
and have enough statistics,
project the events onto the decl. direction,
and produce brightness profiles of the on- and the off-pulse intervals in
the 0.3--5\,keV energy band (PI channels 22--343; Fig.~\ref{fig:fig2}).
The profiles seem to have three components: a sharp peak $|D|<1''$, a relatively broad
shoulder $1''\le |D| \lapp 4''$, and a flat component.
The on-pulse profile (black) has a large core at $|D|<1''$ but matches
well with the off-pulse one (blue) at larger distances.
So we assume that the central peak is the pulsar emission and
the structures at $|D|>1''$ are the PWN. The latter is estimated with a Gaussian function;
we fit the off-pulse profile with a Gaussian plus constant function,
ignoring the central 2$''$ (pulsar component).
The best-fit function is shown in the figure (red dashed).
It is clear that the pulsar's off-pulse emission (blue above red in Fig.~\ref{fig:fig2}) is
very small compared to the PWN ($\sim$5\%).
We also verify that this fraction drops at higher energies.
So using the off-pulse intervals for {\it NuSTAR} PWN studies will be appropriate
because contamination of the pulsar's constant (off-pulse) emission would be very
small ($<$1\%) in the $>$3\,keV band and large apertures $R\ge1'$ used
for {\it NuSTAR} data analyses (Sections~\ref{sec:sec2_4} and \ref{sec:sec2_5}).
	
\begin{figure*}
\centering
\includegraphics[width=7.0 in]{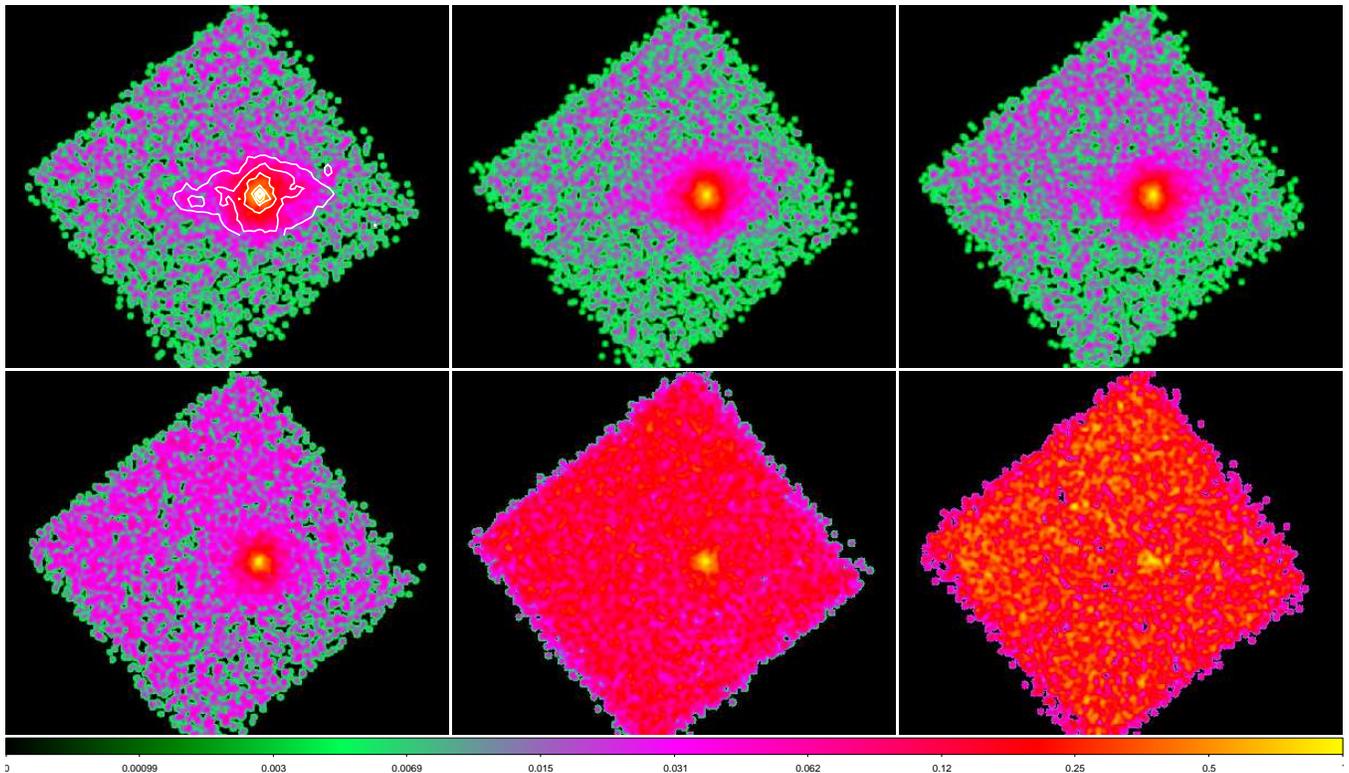}
\figcaption{{\it NuSTAR} images of 3C~58 in six energy bands.
The energy bands for the images are 3--4.5\,keV, 4.5--7\,keV, 7--12\,keV, 12--20\,keV, 20--40\,keV,
and 40--60\,keV from left to right and top to bottom. The images are smoothed and
the scales are adjusted to have a value of 1 at the maximum for better legibility.
{\it Chandra} white contours are overlaid in the top-left panel for reference.
\label{fig:fig3}
}
\vspace{0mm}
\end{figure*}

	Because of the synchrotron burn-off effect,
the size of a young PWN is known to decrease with
energy. This has been seen in some young PWNe \citep[][]{nhra+14,amrk+14,mrha+15},
and here we produce images of the 3C~58 PWN in six energy bands using the {\it NuSTAR} data.
In order to maximize the signal to noise ratio, we combine FPMA/B data together after
aligning them using the central pulsar position.
We extract events in the off-pulse intervals, and make images in six energy
bands.  We show the {\it NuSTAR} images together with {\it Chandra} ACIS contours in Figure~\ref{fig:fig3}.
At low energies, elliptical morphologies similar to that of {\it Chandra} are seen.
The PWN shrinks with increasing energy ; the observed half power diameters (HPDs)
decreases from $120\pm2''$ at 3\,keV to $91\pm1''$ at 14\,keV.
These correspond to $101\pm2''$ and $68\pm2''$ when removing the instrumental PSF \citep[][]{amwb+14} 
in quadrature.
The source is detected up to 60\,keV and the detection
significance is $\sim$5$\sigma$ in the 40--60\,keV band for $R=30''$ extraction.

\subsection{Spatially-resolved Spectral Analysis}
\label{sec:sec2_4}
	The spectrum and its radial variation of 3C~58 in the soft X-ray band ($<10$\,keV)
were previously studied with {\it Chandra} ACIS data by \citet{shvm04}, where the
authors found steepening of the spectrum with distance from the central pulsar.
In that study, the spectrum was measured only out to $100''$ due to the limited field of view
(FoV) and the contamination of a thermal shell \citep[e.g.,][]{bwml+01}.
We extend this to a larger region and higher energies up to $\sim$20\,keV.
Here we perform spectral analyses with the {\it NuSTAR} data and reanalyze the {\it Chandra} data
for comparison with the {\it NuSTAR} results.

\begin{figure}
\centering
\hspace{-5.0 mm}
\includegraphics[width=3.51 in]{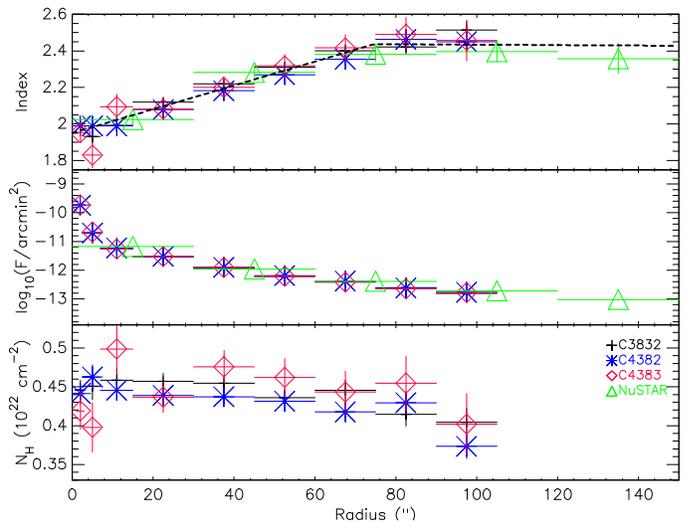}
\figcaption{Radial variation of the PWN spectral parameters measured with {\it NuSTAR} and {\it Chandra}.
Results of {\it Chandra} analyses are denoted in black (ObsID. 3832), blue (ObsID. 4382) and red (ObsID. 4383),
and those of {\it NuSTAR} analyses are shown in green.
{\it Top}: photon index variation. The best-fit broken-line function is shown in a dashed line.
{\it Middle}: variation of the 3--10\,keV brightness.
{\it Bottom}: $N_{\rm H}$ variation measured
with {\it Chandra}.
\label{fig:fig4}
}
\vspace{0mm}
\end{figure}

	We extract {\it NuSTAR} events in the off-pulse intervals using an $R=30''$ circular region
and annular regions with width of $30''$ out to 3$'$ to investigate radial variation of the PWN spectrum.
Background events are extracted in the source-free regions of the detector chips
in which the PWN is detected (mainly detector IDs 0 and 3).
Spectral response files are generated for each region with
the {\tt nuproduct} tool with {\tt extended=yes}.
We group the spectra to have at least 100 events per energy bin and
fit jointly all the spectra in the 3--20\,keV band with absorbed power-law models
having a common absorbing column density ($N_{\rm H}$)
which is held fixed at $4.5\times 10^{21}\ \rm cm^{-2}$.
The results are shown in Figure~\ref{fig:fig4}.
Note that using a broader energy band (e.g., 3--60\,keV) or different background regions
does not change the results because the source dominates in the low-energy band below 20\,keV.

\begin{figure*}
\centering
\begin{tabular}{ccc}
\includegraphics[width=3.35 in]{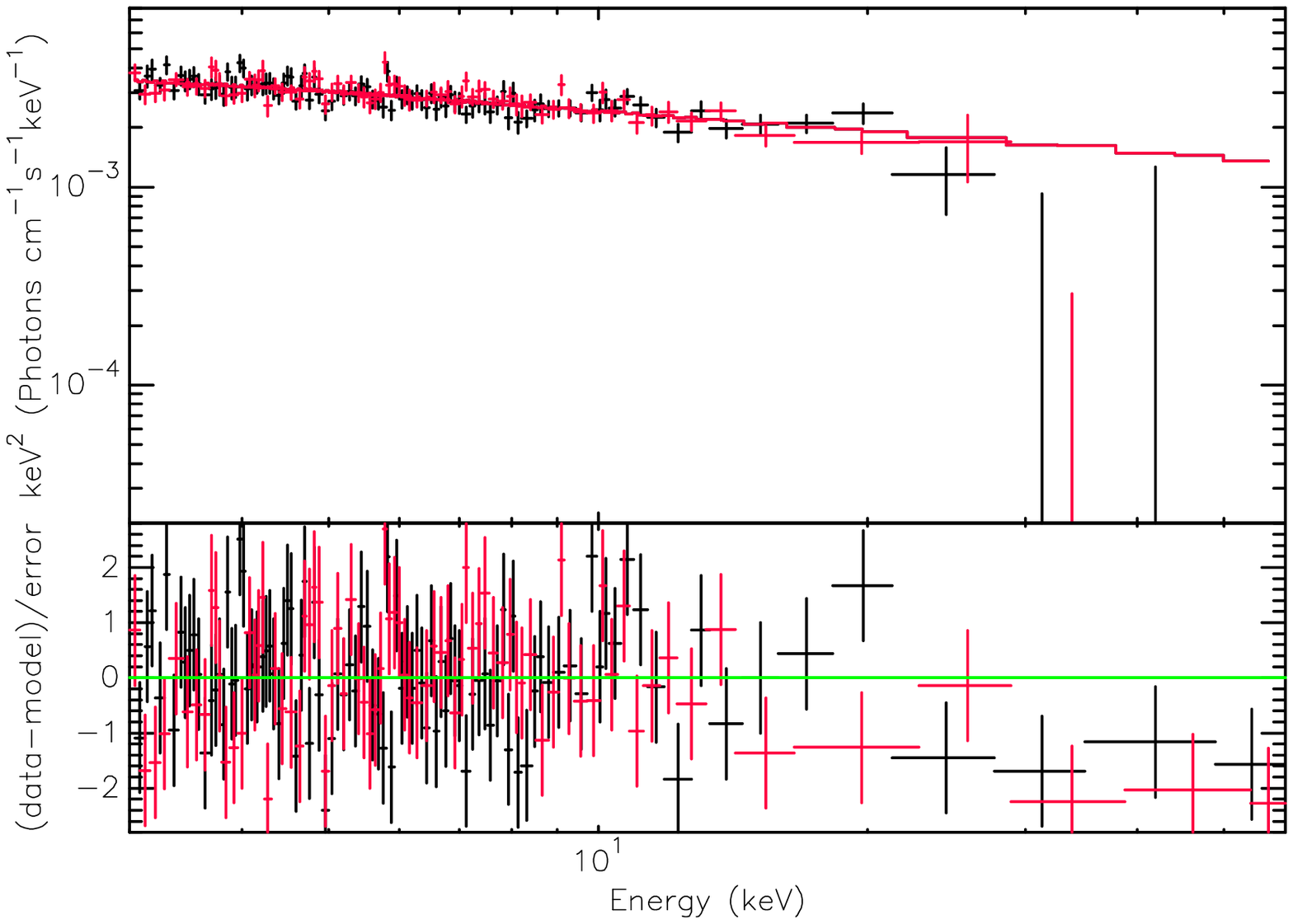} &
\includegraphics[width=3.35 in]{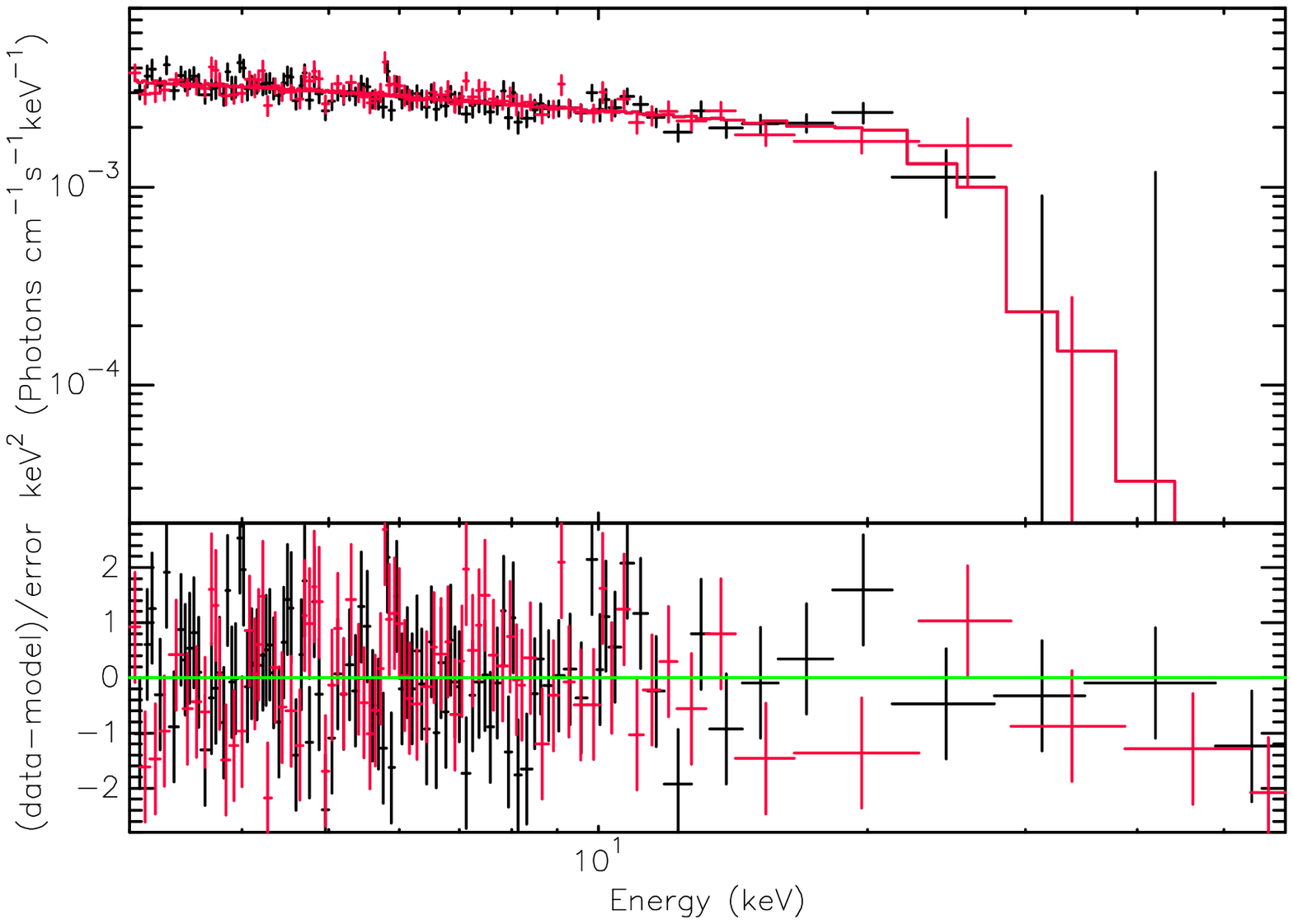} \\
\end{tabular}
\figcaption{Spatially integrated {\it NuSTAR} spectra (3$'$) and residuals (bottom) after fitting.
{\it Left}: power-law fit result.
{\it Right}: broken power-law fit result.
\label{fig:fig5}
}
\vspace{0mm}
\end{figure*}

\newcommand{\marka}{\tablenotemark{a}}
\newcommand{\markb}{\tablenotemark{b}}
\newcommand{\markc}{\tablenotemark{c}}
\newcommand{\markd}{\tablenotemark{d}}
\begin{table*}[t]
\vspace{-0.0in}
\begin{center}
\caption{{\it NuSTAR} fit results for the spatially integrated spectra in the $R=3'$ source region}
\label{ta:ta2}
\vspace{-0.05in}
\scriptsize{
\begin{tabular}{ccccccc} \hline\hline
model\marka & energy range & $\Gamma$       & $E_{\rm b}$\markb &  $\Gamma_2/\alpha$\markc & $F_{\rm X}$\markd & $\chi^2$/dof\\
            &  (keV)       &                & (keV)             &                     &                   &         \\ \hline
PL          & 3--60        &  $2.35\pm0.02$ & $\cdots$          & $\cdots$            & $11.5\pm0.2$     & 303.7/295 \\
BPL         & 3--60        &  $2.33\pm0.02$ & $23.2\pm1.9$      & $9.5\pm8.4$         & $9.0\pm0.3$     & 287.0/293 \\
PLEXP       & 3--60        &  $2.32\pm0.03$ & $30.9\pm5.4$      & $5\pm6$             & $9.1\pm0.4$     & 289.7/293 \\ \hline
\end{tabular}}
\end{center}
\vspace{-0.5 mm}
\footnotesize{
{\bf Notes.} $N_{\rm H}$ is held fixed at $4.5\times10^{21}\ {\rm cm^{-2}}$\\}
$^{\rm a}${PL: power law. BPL: broken power law. PLEXP: power law with a super-exponential cutoff}\\
$^{\rm b}${Break energy for BPL and cut-off energy for PLEXP}\\
$^{\rm c}${Hard power-law index $\Gamma_2$ for BPL and the exponent index $\alpha$ for PLEXP}\\
$^{\rm d}${3--79\,keV flux in units of $10^{-12}\rm\ erg\ cm^{-2}\ s^{-1}$}\\
\end{table*}

	For the {\it Chandra} spectral analysis, we use the ACIS data (Table~\ref{ta:ta1}).
The central pulsar emission is removed using an exclusion region of an $R=1''$ circle. This leaves
$\sim$5\% pulsar contamination in the innermost region, but considering the
very small duty of the pulsation (Fig.~\ref{fig:fig1}) it will be insignificant. We also
check pile-up effects which are only a few \% level just inside the central $1''$ (pulsar region)
and smaller away from the pulsar. Source events are extracted from concentric annular regions with
varying widths (between 2$''$ and 15$''$) out to 100$''$ limited by the detector FoV
for these subarray observations. Backgrounds are extracted from source-free regions.
We generate response files for extended emission, group the spectra to have at least 20 events
per bin, and jointly fit the spectra with absorbed power-law
models in the 0.6--10\,keV band; changing the energy range slightly does not alter the results.
The results are shown in Figure~\ref{fig:fig4}.
The spectral-index and the flux trends agree very well with the {\it NuSTAR}-measured ones.
The spectral index profile (top panel of Fig.~\ref{fig:fig4}) shows a break at $\sim$60--80$''$
similar to that seen in a previous study \citep[][]{shvm04};
our broken-line function fit suggests that the break point is $R_b=75\pm4''$.
We find that the measured $N_{\rm H}$ is slightly lower in the outer regions
(Fig.~\ref{fig:fig4} bottom),
which is probably due to the contamination of a thermal shell as noted by \citet{bwml+01}.

\subsection{Spatially-integrated Spectral Analysis}
\label{sec:sec2_5}

	Next we analyze spatially integrated spectra of the PWN to search for a spectral break
as seen in other young PWNe \citep[e.g., G21.5$-$0.9 and the Crab nebula;][]{nhra+14,mrha+15}.
Because combining {\it Chandra} and {\it NuSTAR} data is not ideal for this study
due to cross-calibration effects \citep[][]{mhma+15}, we focus on the {\it NuSTAR} data.
We extract source events within an $R=3'$ circle in the off-pulse intervals.
Detector background is not uniform across the chips (Fig.~\ref{fig:fig3}), so we extract background
events using circular regions of which area is proportional to the occupancy of the
source region in each detector chip.
We generate response files using the {\tt nuproduct} tool with the {\tt extended=yes} flag,
group the spectra to have at least 100 events
in each spectral bin, and fit the spectra in the 3--60\,keV band with
a power-law model. The model explains the data with a null-hypothesis probability $p\sim0.35$
but leaves a residual trend at high energies (Fig.~\ref{fig:fig5} left).
Motivated by the residual trend (deficit at high energies $>$20\,keV),
we fit the spectra with a broken power-law (BPL) and
a power-law with super-exponential-cutoff models (PLEXP), and find that
these models improve the fit significantly with $f$-test probabilities of
$2.5\times 10^{-4}$ and $10^{-3}$, respectively.
The results are shown in Figure~\ref{fig:fig5} and Table~\ref{ta:ta2}.
We also use different source extraction sizes ($R=1'$ and $2'$) and find that 
the significance (PL vs. BPL) for the break drops
as the size of the source extraction region decreases with
$p\approx 0.2$ and $p\approx 0.04$ for $R=1'$ and $R=2'$
extraction regions, respectively.

	Because background dominates above $\sim$25\,keV, the results may change sensitively
depending on the background selection. We therefore try to use different background regions
for the $R=3'$ source extraction:
(1) extracting background as above (in four chips) but with the size and location
of the background regions changing, and (2) extracting background events in one chip (detector ID 0)
where the largest part of the PWN is detected (Fig.~\ref{fig:fig3}).
Note that the latter is not ideal for background extraction given the
inhomogeneity in background across the detector chips; we do this only for a test.
The significance for the break (PL vs. BPL) changes depending on the
background extraction between $p=4\times10^{-5}$ for case (1) and $p=3\times10^{-2}$ for case (2).

\begin{figure*}[ht]
\centering
\begin{tabular}{cc}
\includegraphics[width=3.5 in]{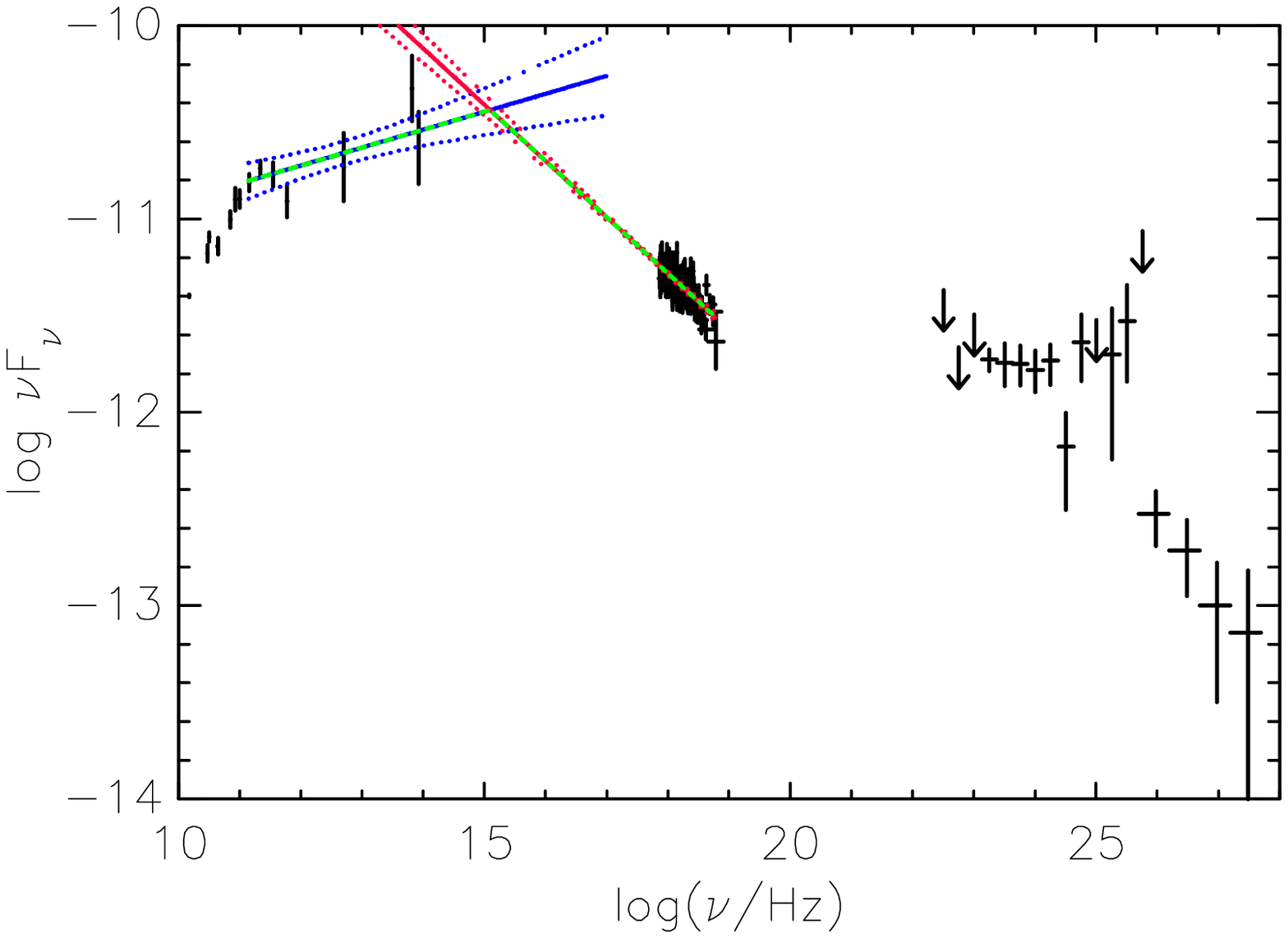} &
\includegraphics[width=3.5 in]{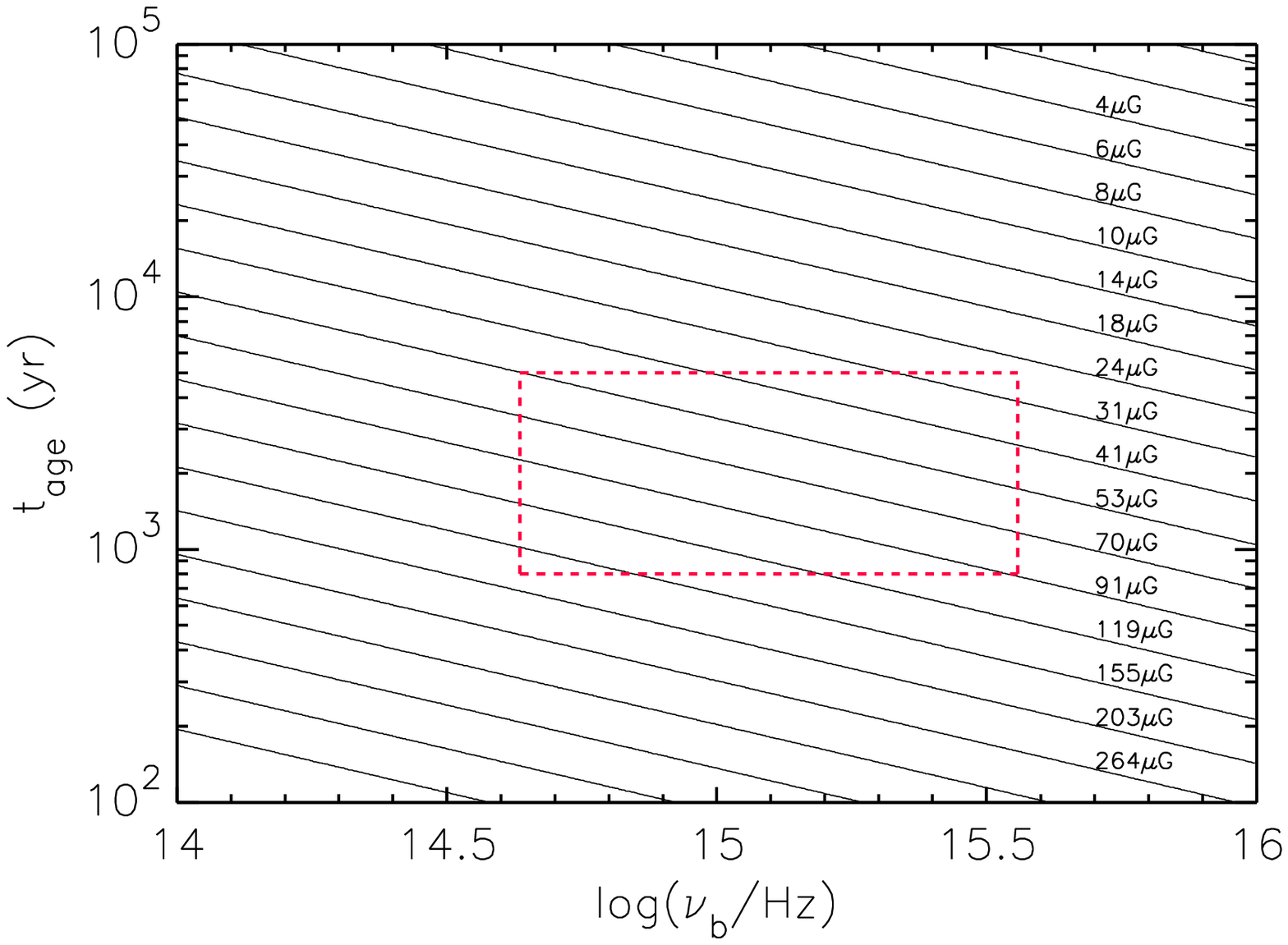} \\
\end{tabular}
\figcaption{{\it Left}: a broadband SED of 3C~58.
The radio-to-IR SED is taken from \citet{shrg+08} and \citet{plancksnr16},
the X-ray SED is measured in this work, and the gamma-ray data are taken
from \citet{3c58magic14} and \citet{ltlg+18}.
Power-law fits with 1-$\sigma$ confidence ranges to the lower-energy
($2\times 10^{11}-10^{14}$\,Hz) and higher-energy ($10^{17}-4\times 10^{18}$\,Hz) data
are shown in blue and red, respectively. A broken power-law fit
to the broadband data ($2\times 10^{11}-4\times 10^{18}$\,Hz) is shown in green.
{\it Right}: relation between the PWN age and the break frequency for various magnetic-field
strengths. The red box shows current estimations on the age and the break frequency.
\label{fig:fig6}
}
\vspace{0mm}
\end{figure*}

	Next, we measure the PWN spectra in a $8'\times 5'$ diamond-shape
region similar to the radio shape \citep[e.g.,][]{shrg+08}.
Because large part of this region lies outside the FoV in the {\it Chandra} observations,
we do not analyze the {\it Chandra} data.
We obtain {\it NuSTAR} spectra and response files using the same method as described
above, fit the spectra with a power-law model, and find
that the spectral index is $\Gamma$ is $2.34\pm0.02$
and the 3--79-keV flux is $11.8\pm0.2\times10^{-12}\rm \ erg\ cm^{-2}\ s^{-1}$
similar to those of the $R=3'$ spectrum.
The high-energy residuals are also visible in these spectra, and a BPL model
is preferred over a simple power law with $p$=0.05--0.0003
depending on the background selection.

	We also analyze the spatially integrated {\it Chandra} spectrum
using $R=2'$ extraction for reference.
In order to minimize contamination from the thermal shell, we ignore the low-energy data
and fit the 3--7\,keV spectra with a power-law model.
We hold $N_{\rm H}$ fixed at the typical value of $4.5\times 10^{21}\ \rm cm^{-2}$, and
measure $\Gamma=2.25\pm0.03$ and $F_{3-10\rm \ keV}=6.14(6)\times10^{-12}\rm \ erg\ cm^{-2}\ s^{-1}$.
Note that the results are not sensitive to the $N_{\rm H}$ value or the energy range.
Although direct comparison with the {\it NuSTAR} results may not be adequate
because of different PSF, we note that the power-law index is similar
to a {\it NuSTAR}-measured one in $R=2'$.

\section{SED in the Radio-to-X-ray band}
\label{sec:sec3}

	With our X-ray and previous measurements \citep[][]{shrg+08,plancksnr16},
we construct a low-energy SED of 3C~58 in the radio to X-ray band.
Figure~\ref{fig:fig6} shows the SED together with high-energy measurements \citep[][]{ltlg+18,3c58magic14}.
For the X-ray data, we use
the {\it NuSTAR} spectrum extracted from the diamond region (Section~\ref{sec:sec2_5})
similar to that used for extracting the radio-to-IR SED.
Note that the hard X-ray SED aligns reasonably well with the previous
{\it Chandra} one \citep[][]{shvm04} which we do not use here because the
measurement does not cover outer regions of the PWN.
The improved X-ray measurement helps to determine accurately the break frequency
between the IR and the X-ray bands which can provide an estimate
of magnetic field or age for the synchrotron-loss dominating PWN \citep[e.g.,][]{tcm13}.
We therefore measure the break frequency by fitting the SED with phenomenological models.

	We fit the IR ($1.4\times 10^{11}<\nu<10^{14}$\,Hz) and the
X-ray data ($10^{17}<\nu<4\times 10^{18}$\,Hz) separately with power-law functions.
The frequency ranges are chosen to avoid the radio ($\nu\sim10^{11}$\,Hz) and the X-ray breaks
($\sim 5\times 10^{18}$\,Hz) which are assumed
to be intrinsic to the electron distribution \citep[e.g.,][]{baa11}.
We extend the best-fit functions to find the frequency
at which the two power-law functions cross (blue and red lines in Fig.~\ref{fig:fig6} left).
We also fit the SED ($1.4\times 10^{11}<\nu<6\times 10^{18}$\,Hz) with a broken power-law function (green)
and find that the break frequency is between $4.3\times 10^{14}$\,Hz and $3.6\times 10^{15}$\,Hz
and the degree of break is $\Delta\Gamma=0.38\pm0.05$
with 1-$\sigma$ confidence. Note that varying the frequency range for the fit or the {\it NuSTAR}
flux \citep[e.g., calibration uncertainty;][]{mhma+15} does not alter the results significantly.

	For a synchrotron-loss dominated PWN with homogeneous magnetic field,
there is a simple relation among the frequency of the cooling break ($\nu_b$),
age of the PWN ($t_{\rm age}$), and the magnetic field strength ($B$):
$t_{\rm age}=4.12\times 10^4 (\frac{B}{1G})^{-1.5} (\frac{\nu_{\rm b}}{1Hz})^{-0.5}$\,years \citep[e.g.,][]{shrg+08}.
This relation can give us an estimate of one quantity if the other two are known.
Figure~\ref{fig:fig6} shows the relations between the age and the break frequency for
various magnetic-field strengths. Our measurement of the break frequency and previous
estimation of the age can be used to constrain the magnetic-field strength in 3C~58.
The red box in the figure shows the previously-suggested parameter space
for $t_{\rm age}$ and $\nu_{b}$ for 3C~58, and from this we infer $B$ to be between 30$\mu$G and 200$\mu$G.

\section{Discussion and Conclusions}
\label{sec:sec4}
	We analyzed new {\it NuSTAR} and archival {\it Chandra} data to measure the X-ray
spectrum of 3C~58 and search for a spectral break in the X-ray band. We carefully selected off-pulse
intervals so that the pulsar emission is ignorable in the {\it NuSTAR} data analyses.
The size of the PWN decreases with increasing energy, which we attribute to the 
synchrotron burn-off effect. The radial profile of the spectral index breaks
at $R\approx 80''$, and the spatially integrated X-ray spectrum of 3C~58 shows a hint of
a spectral break at $\approx25$\,keV.

	A radial change (out to $\sim$100$''$) of the 2.2--8\,keV spectral index was measured
for 3C~58 \citep[][]{shvm04} previously.  From this measurement,
\citet{tc12} inferred a maximum electron energy of 40\,TeV using a diffusion model
for an assumed equipartition magnetic-field strength of 80\,G \citep{gs92}.
Our studies made with {\it NuSTAR}
revealed a similar radial trend of the X-ray spectrum to
that measured with {\it Chandra} \citep[][]{shvm04}, and extended the measurements to higher energies
$\sim$20\,keV and to a larger distance (Fig.~\ref{fig:fig4}).
This then will increase the inferred maximum electron energy because
electrons should be able to emit higher-energy photons for the same magnetic field.
Using analytic formulae \citep[equations 8 and 14 of][]{tc12} and our measurements,
we calculate the maximum electron energy to be $\sim$200\,TeV;
this approximate calculation provides a rough estimate, but actual simulations
need to be performed to improve the estimate.
Note that we used $R_{\rm PWN}=2'$ \citep[][]{bwml+01}
in our calculation even though the properties are measured out to R$\sim$3$'$.
This is because photons in the outermost region in the {\it NuSTAR} data are probably
produced by PSF scattering.

	The low-energy SED in the radio to X-ray band (Fig.~\ref{fig:fig6} left) suggests
a spectral (cooling) break $\Delta\Gamma\sim0.4$ at $\sim$$10^{15}$\,Hz.
From the measured frequency range for the cooling break and the assumed ages of 3C~58,
we inferred the magnetic-field strength (assuming constant field over the PWN)
to be $\sim$30--200$\mu$\,G.
This observationally-estimated $B$ agrees with the previous estimations based on SED modelings
\citep[][]{tt13,tcm13} or $\sim$40\,$\mu$G estimated with MHD simulations \citep[][]{pvle16}.
Note that \citet{tt13} and \citet{tcm13} used the cooling break in their SED models,
so the agreement was anticipated.
While it is difficult to estimate both $B$ and $t_{age}$ simultaneously because of parameter degeneracy,
a measurement of one parameter (e.g., magnetic field strength using radio polarization)
will be very helpful to infer the other and to confirm whether or not the PWN
is associated with SN~1181. Of course, the IR and the X-ray SED measurements
also need to be improved in order to determine the frequency of the cooling break more precisely.

	We find a hint of a break in the spatially-integrated X-ray spectrum of 3C~58 at
$\sim$25\,keV. The detection significance is relatively high
for a larger extraction region, but the significance drops as the extraction region becomes smaller.
This is perhaps because of decreasing statistics for smaller extraction regions.
Although the detection significance is not very high and
the degree of the break ($\Delta \Gamma$) is not well measured,
we speculate on a possible reason.
In synchrotron emission models for PWNe \citep[e.g.,][]{kc84a,tc12},
the synchrotron cooling pushes the high-energy cutoff in the electron distribution
to low energies with time (i.e., distance from the pulsar),
and so the integrated photon spectrum of a PWN is expected to be a simple power law
in the X-ray band up to a frequency at which the maximum-energy electrons radiate.
In these models, the 25-keV spectral break in
the photon spectrum corresponds to the cutoff energy of the electron distribution,
implying a maximum electron energy of $\sim$140\,TeV for an
assumed magnetic-field strength of 80$\mu$\,G. Note that this is larger than
40\,TeV inferred from a previous diffusion modeling \citep[][]{tc12}.
Although detailed simulations with diffusion and particle transport are
needed for a better estimation of the maximum electron energy,
our estimation is of the same order of magnitude of previous works.
Nevertheless, if the 25-keV break is real, it may imply that the injected electron
distribution cuts off at a lower energy than assumed in
SED models for 3C~58 \citep[][]{tt13,tcm13,ltlg+18}. So the models may need
some modifications. The effect of the break
is to change synchrotron and IC SEDs at the highest energies,
but the changes are expected to be very small because of
decreasing flux with energy (synchrotron) and the Klein-Nishina suppression (IC).

	In a recent model \citep[][]{ltks+18}
existence of two populations of particles, accelerated by shock and magnetic
reconnection in PWNe, are suggested. A spectrum of the latter population may extend
to higher energies and make a small step in the spectrum when combined with the former.
A hint of a spectral step is suggested in the Crab nebula at $\sim$130\,keV \citep[][]{mhz10,ltks+18}.
So it will be interesting to see whether the spectrum of 3C~58 keeps curving down
$\ge$25\,keV (i.e., cutoff), or the break is actually a small step
and another power-law spectrum appears at higher energies;
the latter may imply another population of particles, i.e., acceleration mechanism in
PWNe \citep[][]{ltks+18}.
This can be tested in the hard X-ray and/or the soft gamma-ray bands
with deeper {\it NuSTAR} and future {\it AMEGO} \citep[][]{amego17} observations.

	Our measurements of X-ray properties of 3C~58 add more information
to the acceleration/emission models of PWNe. While current PWN models
are able to reproduce the observed SED or the radial profile of the soft-X-ray
spectral index, it is not clear whether the models can accommodate the hard X-ray
properties, in particular, the possible spectral break. So confirming or disproving
the existence of the break observationally is crucial.
With the well measured broadband SED and the possible X-ray break,
3C~58 may give us new insights into particle acceleration and flow in PWNe.

\bigskip
\bigskip

\acknowledgments

We thank the anonymous referee for the careful reading of the paper and
insightful comments. This research was supported by Basic Science Research Program through
the National Research Foundation of Korea (NRF)
funded by the Ministry of Science, ICT \& Future Planning (NRF-2017R1C1B2004566).

\vspace{5mm}
\facilities{CXO, NuSTAR}
\software{HEAsoft (v6.24; HEASARC 2014), CIAO \citep[v4.10; ][]{fmab+06}, XSPEC \citep[][]{a96}}

\bibliographystyle{apj}
\bibliography{GBINARY,BLLacs,PSRBINARY,PWN,STATISTICS,FERMIBASE,COMPUTING,INSTRUMENT,ABSORB}

\end{document}